# CONTROLLER SYNTHESIS METHOD FOR MULTI-AGENT SYSTEM BASED ON TEMPORAL LOGIC SPECIFICATION

Ruohan Huang[1,2,3] and Zining Cao[1,2,3*]

[1] College of Computer Science and Technology, Nanjing University of Aeronautics and Astronautics, Nanjing 211106, P. R. China
[2] Ministry Key Laboratory for Safety-Critical Software Development and Verification, Nanjing University of Aeronautics and Astronautics, Nanjing 211106, P. R. China
[3] Collaborative Innovation Center of Novel Software Technology and Industrialization, Nanjing 210023, P.R. China
*Corresponding author. Email: caozn@nuaa.edu.cn

## ABSTRACT

*Controller synthesis is a theoretical approach to the systematic design of discrete event systems. It constructs a controller to provide feedback and control to the system, ensuring it meets specified control specifications. Traditional controller synthesis methods often use formal languages to describe control specifications and are mainly oriented towards single-agent and non-probabilistic systems. With the increasing complexity of systems, the control requirements that need to be satisfied also become more complex. Based on this, this paper proposes a controller synthesis method for semi-cooperative semi-competitive multi-agent probabilistic discrete event systems to solve the controller synthesis problem based on temporal logic specifications. The controller can ensure the satisfaction of specifications to a certain extent. The specification is given in the form of a linear temporal logic formula. This paper designs a controller synthesis algorithm that combines probabilistic model checking. Finally, the effectiveness of this method is verified through a case study.*

## 1. INTRODUCTION

Discrete-event systems (DES) [1] are discrete-state, event-driven systems whose state evolution is determined exclusively by the occurrence of discrete events: a state transition takes place if and only if a discrete event occurs, i.e., the interaction of discrete events drives the change of system state. With the development of science and technology, DES have permeated nearly every sector of social life. Representative DES in modern society include online shopping platforms, communication networks, traffic control systems, flexible manufacturing systems, and automated logistics systems. Consequently, research on DES is of significant theoretical and practical value. The study of DES mainly addresses two categories: performance analysis and controller synthesis. This paper focuses on the latter.

Controller synthesis originates from control theory. The controller synthesis problem for DES was first formulated by Ramadge and Wonham [2]. Controller synthesis provides a declarative-specification-based design methodology for DES. Its primary task is to synthesize a controller that, when placed in closed-loop configuration with the plant, enforces—over time—that the system behaviour satisfies a given specification [3]. In [2], the Ramadge–Wonham (R–W) framework, based on finite-state automata and formal languages, was proposed, laying the foundation for DES controller synthesis. The authors synthesized controllers for deterministic

DES by computing the supremal controllable sublanguage of an automaton. Building on this work, [4] extended the approach to probabilistic DES and probabilistic languages, providing a controller synthesis algorithm via a fixed-point iteration scheme.

Nevertheless, the above studies exhibit three principal limitations: (i) they address only monolithic systems, neglecting multi-agent systems—especially those of a semi-cooperative and semi-competitive nature—which narrows their applicability; (ii) they consider exclusively systems whose behaviours terminate, leaving non-terminating behaviours unexplored; and (iii) the desired control properties are specified by formal languages, yet it is generally difficult to translate such properties into formal languages precisely, and formal languages become inadequate when specifications grow complex. To overcome these drawbacks, temporal logics have been introduced into controller synthesis. In [5], it proposes a controller synthesis method based on computation tree logic (CTL). In [6–9], they investigate linear temporal logic (LTL) and present synthesis techniques founded on reactive synthesis [10], reinforcement learning [11], and bounded synthesis [12], respectively.

Although the above works study temporal-logic-based controller synthesis, they concentrate primarily on deterministic DES and single-agent system. Consequently, controller synthesis for multi-agent probabilistic DES with respect to LTL specifications remains an open problem. To address the limitations of existing research, this paper proposes a controller synthesis methodology for multi-agent probabilistic DES subject to LTL specifications. The main contributions are as follows:

(1) A multi-agent probabilistic transition system model is introduced to capture semi-cooperative and semi-competitive multi-agent probabilistic DES.

(2) Leveraging probabilistic model checking techniques, a controller synthesis approach based on probabilistic model checking is devised.

(3) The effectiveness of the proposed method is validated through a case study that demonstrates its potential in controller synthesis.

The structure of this paper is as follows: Section 2 reviews the basic concepts. Section 3 gives the definition of multi-agent probabilistic transition system and controller. Section 4 introduces the formal definition of the research problem and the algorithm for controller synthesis in detail. Section 5 presents a case study. Section 6 concludes with a summary and directions for future research.

## 2. BASIC CONCEPTS

### 2.1. Transition System

Transition systems (TS) [13] are widely applied to model discrete state systems. The definition of a TS is as follows.

Definition 1: A TS can be represented as a four tuple, i.e., $T =< X, A, f, x_0 >$, where
(1) $X$ is the set of states.
(2) $A$ is the set of actions.
(3) $f : X \times A \rightarrow X$ is the transition function.
(4) $x_0$ is the initial state.

An infinite sequence $\rho = x_0 a_0 x_1 a_1 x_2 ... \in X(AX)^\infty$ is called a run (or infinite path) of $T$. A finite sequence $h = x_0 a_0 x_1 ... a_n x_n \in X(AX)^*$ is called a history (or finite path) of $T$. A finite or infinite path is said to be *initialized* if its first element is the initial state $x_0$. We denote by $Runs(T)$ the set of all initialized infinite paths of $T$, and by $His(T)$ the set of all initialized finite paths of $T$.

## 2.2. Linear Temporal Logic

Linear temporal logic (LTL) is obtained by adding temporal operators to propositional logic. This paper only provides a brief introduction to the syntax and semantics of LTL, and for detailed information about LTL, one can refer to the literature [13].

Definition 2: The LTL formulas acting on the set of atomic propositions $AP$ are formed according to the following syntax:

$$\varphi ::= true \mid p \mid \neg\varphi \mid \varphi_1 \wedge \varphi_2 \mid X\varphi \mid \varphi_1 U \varphi_2$$

where $p \in AP$; $\varphi$, $\varphi_1$, $\varphi_2$ are LTL formulas; $X \cdot$, $\cdot U \cdot$ are temporal operators.

Other Boolean connectives such as $\vee$, $\rightarrow$, $\leftrightarrow$ and temporal operators such as $F \cdot$, $G \cdot$ can be derived as follows:

$$\varphi_1 \vee \varphi_2 = \neg(\neg\varphi_1 \wedge \neg\varphi_2)$$
$$\varphi_1 \rightarrow \varphi_2 = \neg\varphi_1 \vee \varphi_2$$
$$\varphi_1 \leftrightarrow \varphi_2 = (\varphi_1 \rightarrow \varphi_2) \wedge (\varphi_2 \rightarrow \varphi_1)$$
$$F\varphi = true U \varphi$$
$$G\varphi = \neg F \neg \varphi$$

Let $AP$ be a set of atomic propositions. An infinite sequence over the power set $2^{AP}$ is referred to as an infinite word.

Definition 3: The semantics of LTL are defined over infinite words. Given an infinite word $\omega = \alpha_0 \alpha_1 \ldots \in (2^{AP})^\infty$, let $\omega^i = \alpha_i \alpha_{i+1} \ldots$ denotes the suffix of $\omega$ starting at $\alpha_i$. Therefore,

(1) $\omega \vDash true$.
(2) $\omega \vDash p$, if and only if $p \in \alpha_0$.
(3) $\omega \vDash \neg\varphi$, if and only if $\omega \nvDash \varphi$.
(4) $\omega \vDash \varphi_1 \wedge \varphi_2$, if and only if $\omega \vDash \varphi_1$ and $\omega \vDash \varphi_2$.
(5) $\omega \vDash X\varphi$, if and only if $\omega^1 \vDash \varphi$.
(6) $\omega \vDash \varphi_1 U \varphi_2$, if and only if there exist $j \in N$ such that $\omega^j \vDash \varphi_2$, and for all $0 \leq i < j$, $\omega^i \vDash \varphi_1$.

## 2.3. Deterministic Rabin Automaton

It is well known that any LTL formula can be transformed into an equivalent automaton, in this paper, we use deterministic Rabin automaton (DRA) to represent LTL formulas, and the details of the construction can be found in [14].

Definition 4: A DRA is a five tuple, denoted by $R = <X, I, f, x_0, Acc>$, where

(1) $X$ is the set of states.
(2) $I$ is the input alphabet. Typically, $I = 2^{AP}$, where $AP$ is the set of atomic propositions.
(3) $f : X \times I \rightarrow X$ is the transition function.
(4) $x_0$ is the initial state.
(5) $Acc = \{(L_1, K_1), (L_2, K_2), \ldots, (L_d, K_d)\}$ is the Rabin acceptance condition, where $(L_i, K_i) \in 2^X \times 2^X$ are Rabin pairs, $d \geq 1$ and $d \in N$.

Let $\omega = \alpha_0 \alpha_1 \ldots \in I^\infty$ be an infinite word. An infinite sequence $\rho = x_0 \alpha_0 x_1 \alpha_1 \ldots \in X(IX)^\infty$ induced by $\omega$ is called a run (or infinite path) of $R$. The infinite word $\omega$ is accepted by $R$ if there exists an index $1 \leq i \leq d$ such that $Inf(\rho) \cap L_i = \varnothing \wedge Inf(\rho) \cap K_i \neq \varnothing$, where $Inf(\rho) \subseteq X$ denotes the set of states visited infinitely often along $\rho$. The set of all infinite words over the alphabet $I$ accepted by $R$ is called the language of $R$ and is denoted by $\mathcal{L}(R)$.

For every LTL formula $\varphi$, one can construct a DRA $R_\varphi$ over the input alphabet $I = 2^{AP}$ such that for every infinite word $\omega = \alpha_0 \alpha_1 \ldots \in (2^{AP})^\infty$, having $\omega \vDash \varphi \Leftrightarrow \omega \in \mathcal{L}(R_\varphi)$.

### 2.4. Probability Measure

Reasoning about quantitative properties of probabilistic models relies on measure theory, particularly on the notions of probability spaces and $\sigma$-algebras. In what follows, we provide only a concise introduction to the concepts required in this paper, comprehensive details can be found in [15].

Definition 5: A $\sigma$-algebra can be defined as a tuple $(\Omega, \Lambda)$, where $\Omega$ is a non-empty set, $\Lambda \subseteq 2^\Omega$ is a set of subsets of $\Omega$, and satisfying the following conditions:
(1) $\varnothing \in \Lambda$, $\Omega \in \Lambda$.
(2) If $E \in \Lambda$, then its complement $\Omega \setminus E \in \Lambda$, i.e., $\Lambda$ is closed under complementation.
(3) If for $i \in N$, $E_0, E_1, \ldots \in \Lambda$, then countable union $\bigcup_{i=0}^\infty E_i \in \Lambda$ and countable intersection $\bigcap_{i=0}^\infty E_i \in \Lambda$, i.e., $\Lambda$ is closed under countable set operations.

The elements of $\Omega$ are often called *outcomes*, and the elements of $\Lambda$ are called *events*. According to the literature [15], for any set $\Omega$ and $\Theta \subseteq 2^\Omega$, there exists a unique smallest $\sigma$-algebra containing $\Theta$. The $\sigma$-algebra generated by $\Theta$ is denoted by $\Lambda_\Theta$, and $\Theta$ is the *basic event* in $\Lambda_\Theta$.

Definition 6: A probability measure on $(\Omega, \Lambda)$ is a function $Pr: \Lambda \to [0,1]$ that satisfies the following conditions:
(1) $Pr(\varnothing) = 0$.
(2) $Pr(\Omega) = 1$.
(3) For any sequence of pairwise disjoint events $E_0, E_1, \ldots \in \Lambda$, it holds that $Pr(\bigcup_{i=0}^\infty E_i) = \sum_{i=0}^\infty Pr(E_i)$.

Definition 7: A probability space is a triple $(\Omega, \Lambda, Pr)$, where $(\Omega, \Lambda)$ is a $\sigma$-algebra and $Pr$ is a probability measure defined on $(\Omega, \Lambda)$.

## 3. MULTI-AGENT PROBABILISTIC TRANSITION SYSTEM

### 3.1. System Model

To model multi-agent probabilistic discrete event systems, this paper extends the transition system introduced in Section 2.1 with both multi-agent and probabilistic features, yielding the Multi-Agent Probabilistic Transition System (MPTS). The MPTS captures the concurrent and probabilistic interactions of multiple players (or agents) whose joint actions influence the state transitions of the system. The formal definition of an MPTS is given below.

Definition 8: An MPTS is formally represented by a nine tuple, i.e.,
$$M = <\Sigma, X, A, \{p_1, \ldots, p_n\}, p, f, x_0, AP, L>$$
where
(1) $\Sigma = \{1, 2, \ldots, k\}$ is the set of players. This set is partitioned into two disjoint subsets $\Sigma^P$ and $\Sigma^Q$, where $\Sigma^P$ contains the cooperative players, and $\Sigma^Q = \Sigma \setminus \Sigma^P$ contains the players that are in competition with those in $\Sigma^P$.
(2) $X$ is the set of states.

(3) $A$ is the set of actions. Let $\vec{A} = \{\vec{a} = <a_1,...,a_k> | \forall i \in \Sigma, a_i \in A\}$ denote the set of joint actions. We write $\vec{A^P}$ for the joint actions of players in $\Sigma^P$ and $\vec{A^Q}$ for the joint actions of players in $\Sigma^Q$. Clearly, $\vec{A} = \vec{A^P} \times \vec{A^Q}$.

(4) $\{p_1,...,p_n\}$ is a family of probability distributions, where $n = k$. For every player $i \in \Sigma$, the function $p_i : X \times A \to [0,1]$ gives the probability that player $i$ executes an action in a state. For non-terminating MPTSs, i.e., those in which at least one joint action is enabled in every state, the equality $\sum_{a \in A} p_i(x,a) = 1$ holds for every state $x \in X$. Unless explicitly stated otherwise, all MPTSs discussed in this paper are assumed to be non-terminating.

(5) $p : X \times \vec{A} \to [0,1]$ denotes the probability of executing each joint action in the current state. For every state $x \in X$, the probability of executing joint action $\vec{a} = <a_1,...,a_k> \in \vec{A}$ is $p(x,\vec{a}) = \prod_{i \in \Sigma} p_i(x,a_i)$, and it satisfies $\sum_{\vec{a} \in \vec{A}(x)} p(x,\vec{a}) = 1$, where $\vec{A}(x)$ denotes the set of joint actions that can be enabled in state $x$.

(6) $f : X \times \vec{A} \to X$ is the transition function.

(7) $x_0 \in X$ is the initial state.

(8) $AP$ is the set of atomic propositions.

(9) $L : X \to 2^{AP}$ is the label function.

A finite sequence $h = x_0 \vec{a_0} x_1 ... \vec{a_n} x_n \in X(\vec{A}X)^*$ is called a history (or finite path) of $M$. An infinite sequence $\rho = x_0 \vec{a_0} x_1 \vec{a_1} x_2 ... \in X(\vec{A}X)^\infty$ is called a run (or infinite path) of $M$. For a history or run, if its first element is the initial state $x_0$, it is said to be initialized. The sets of all initialized histories and runs of $M$ are denoted by $His(M)$ and $Runs(M)$, respectively.

Each run generated by $M$ can be associated with a unique infinite word through the label function. This infinite word can be expressed using the extended label function, i.e., for a run $\rho = x_0 \vec{a_0} x_1 \vec{a_1} x_2 ... \in X(\vec{A}X)^\infty$, we have $\omega = L(\rho) = L(x_0)L(x_1)L(x_2)... \in (2^{AP})^\infty$. For a given run $\rho \in Runs(M)$ of $M$ and an LTL formula $\varphi$, it holds that $M, \rho \vDash \varphi$ if and only if $L(\rho) \vDash \varphi$.

### 3.2. Controller

A controller is a control element or strategy that operates on a system model, ensuring that the system's behaviour adheres to specified control specifications by restricting the executable actions in each state.

Definition 9: Given an MPTS $M$, a controller acting on $M$ can be formally defined as a mapping function, i.e.,

$$C : X \to \vec{A^P}$$

This function maps a state in $M$ to a joint action in $\vec{A^P}$. Intuitively, the controller only restricts the action choices of the players in $\Sigma^P$, without imposing constraints on the actions of the players in $\Sigma^Q$.

### 3.3. Controlled System Model

Given a system model $M = <\Sigma, X, A, \{p_1,...,p_n\}, p, f, x_0, AP, L>$ and a controller $C : X \to \vec{A^P}$, the controlled system is denoted by $C/M$. Essentially, the controlled system is a sub model of the system model, where the transitions between states in the controlled system are determined by the controller $C$.

Definition 10: A controlled system can be represented by an MPTS, denoted as

$$C/M = <\Sigma^{C/M}, X^{C/M}, A^{C/M}, \{p_1,...,p_n\}^{C/M}, p^{C/M}, f^{C/M}, x_0^{C/M}, AP^{C/M}, L^{C/M}>$$

where

(1) $\Sigma^{C/M}$ denotes the set of players, satisfying $\Sigma^{C/M} = \Sigma$, meaning that the set of players in the controlled system is consistent with that of the system model $M$.

(2) $X^{C/M}$ denotes the set of states, satisfying $X^{C/M} = X$, meaning that the state set of the controlled system is consistent with that of the system model $M$.

(3) $A^{C/M}$ denotes the set of actions, satisfying $A^{C/M} = A$, meaning that the action set of the controlled system is consistent with that of the system model $M$. In the controlled system, the set of joint actions $\vec{A}^{C/M} = \{<\vec{a^P}, \vec{a^Q}> | \forall x \in X^{C/M}, \vec{a^P} = C(x), \vec{a^Q} \in \vec{A^Q}(x)\}$, where $\vec{a^P} = C(x)$ represents the joint action that players in $\Sigma^P$ can execute under the control of $C$ in state $x$, and $\vec{a^Q} \in \vec{A^Q}(x)$ represents the joint action that players in $\Sigma^Q$ can execute in state $x$.

(4) $\{p_1^{C/M},...,p_n^{C/M}\}$ denotes the set of probability distributions, satisfying $\{p_1^{C/M},...,p_n^{C/M}\} = \{p_1,...,p_n\}$, meaning that the probability distributions of each player in the controlled system are consistent with that of the system model $M$.

(5) $p^{C/M}: X^{C/M} \times \vec{A}^{C/M} \to [0,1]$ denotes the probability function for state-action pairs. For any state $x \in X^{C/M}$, the probability of executing the joint action $\vec{a} = <\vec{a^P}, \vec{a^Q}> \in \vec{A}^{C/M}$ is given by $p^{C/M}(x, \vec{a}) = \prod_{i \in \Sigma^Q} p_i^{C/M}(x, a_i)$, and it satisfies $\sum_{\vec{a} \in \vec{A}^{C/M}(x)} p^{C/M}(x, \vec{a}) = 1$, where $\vec{A}^{C/M}(x) = C(x) \times \vec{A^Q}(x)$ representing the set of joint actions enabled by the controller $C$ in state $x$.

(6) $f^{C/M}: X^{C/M} \times \vec{A}^{C/M} \to X^{C/M}$ denotes the transition function. For any state $x \in X^{C/M}$ and joint action $\vec{a} = <\vec{a^P}, \vec{a^Q}> \in \vec{A}^{C/M}$, satisfying $f^{C/M}(x, \vec{a}) = f(x, \vec{a})$, meaning that the controlled system reaches the same successor state as the system model $M$ when executing the joint action $\vec{a}$ in state $x$, where $f$ is the transition function of $M$.

(7) $x_0^{C/M}$ denotes the initial state, satisfying $x_0^{C/M} = x_0$, meaning that the initial state of the controlled system is consistent with that of the system model $M$.

(8) $AP^{C/M}$ denotes the set of atomic propositions, satisfying $AP^{C/M} = AP$, meaning that the set of atomic propositions in the controlled system is consistent with that of the system model $M$.

(9) $L^{C/M}: X^{C/M} \to 2^{AP^{C/M}}$ denotes the label function. For any state $x \in X^{C/M}$, it satisfies $L^{C/M}(x) = L(x)$, meaning that the labeling of each state in the controlled system is consistent with that of the system model $M$.

The sets of initialized runs and histories of the controlled system are denoted by $Runs(C/M)$ and $His(C/M)$, respectively. There is a relationship between the runs of the controlled system and the runs of the system model: for any $\rho = x_0 <\vec{a_0^P}, \vec{a_0^Q}> x_1... \in Runs(M)$, if and only if for every $j \in N$, $\vec{a_j^P} = C(x_j)$, then $\rho \in Runs(C/M)$, where $\vec{a_j^P}$ represents the joint action formed by the actions of players in $\Sigma^P$, and $\vec{a_j^Q}$ represents the joint action formed by the actions of players in $\Sigma^Q$.

### 3.4. Probability Measure on Path Sets

To conduct a probabilistic analysis of the given model, this section establishes a correspondence between the concepts described in Section 2.4 and the path sets of MPTS, and provides the probability measure of MPTS path sets along with relevant definitions.

Given an MPTS $M$, the infinite paths in $M$ correspond to the outcomes in $\Omega$, hence $\Omega^M = Runs(M)$. The $\sigma$-algebra associated with the path sets of $M$ is generated by the cylinder sets derived from the finite paths in $M$.

Definition 11: The cylinder set of a finite path $h = x_0\vec{a_0}...\vec{a_n}x_n \in His(M)$ is defined as
$$Cyl(h) = \{\rho \in Runs(M) \mid h \in pref(\rho)\}$$
where $\rho$ denotes an infinite path, and $pref(\rho)$ denotes the set of finite prefixes of the infinite path $\rho$.

Therefore, the cylinder set obtained by extending the finite path $h$ includes all infinite paths with $h$ as their finite prefix. The cylinder set is equivalent to the basic event in the $\sigma$-algebra associated with the path set of $M$.

Definition 12: The $\sigma$-algebra associated with the path set of $M$ is the smallest $\sigma$-algebra containing all cylinder sets, denoted as $\Lambda^M$.

From the classical concepts of measure theory, there exists a probability measure $Pr^M$ on the $\sigma$-algebra $\Lambda^M$ of the path set of $M$, where the probability of the cylinder set is given by
$$Pr^M(Cyl(x_0\vec{a_0}...\vec{a_n}x_n)) = \prod_{0 \leq i < n} p(x_i, \vec{a_i})$$
For $Cyl(x_0)$, the $Pr^M(Cyl(x_0))$ is defined as 1.

Based on this, a probability measure on MPTS can be defined. This measure allows us to study the probability that a controller $C$ acting on $M$ satisfies an LTL formula $\varphi$.

Definition 13: Given an MPTS $M$ and an LTL formula $\varphi$, the probability that $M$ satisfies $\varphi$ is
$$Pr^M(\varphi) = Pr^M\{\rho \in Runs(M) \mid M, \rho \vDash \varphi\}$$

## 4. CONTROLLER SYNTHESIS

### 4.1. Problem Statement

The main research problem of this paper is to synthesize controllers for multi-agent probabilistic discrete event systems with semi-cooperative and semi-competitive relationships, with the system's control specifications given by LTL formulas. The problem is formally stated as follows.

Problem 1: Given a system model $M$, an LTL formula $\varphi$, and a probability threshold $\vartheta \in [0,1]$, find a controller $C$ such that $Pr^{C/M}(\varphi) = Pr^{C/M}\{\rho \in Runs(C/M) \mid C/M, \rho \vDash \varphi\} \geq \vartheta$.

To solve this problem, Algorithm 1 provides a detailed process description. In this algorithm, the LTL formula is first translated into a DRA. Then, all possible controllers of the system model are enumerated, and corresponding controlled systems are generated based on these controllers. Next, a product automaton is constructed by using the controlled system and the DRA, and the accepting component set is calculated based on the product automaton. Finally, the probability of reaching the accepting component set is computed by using the probability model checking. If the computed probability meets the given threshold, the corresponding controller is returned; otherwise, all possible controllers are traversed, and the process is repeated until a controller satisfying the requirements is found or all controllers have been tried, in which case a null value is returned.

**Algorithm 1**: *ControllerSynthesis*($M, \varphi, \vartheta$)

**Input**: System Model $M$, LTL Formula $\varphi$, Probability Threshold $\vartheta$.

**Output**: Controller $C$.

1:    $R := ConvertLTLToDRA(\varphi)$; //Convert $\varphi$ into DRA $R$.
2:    $Cs := EnumController(M)$; //Enumerate all possible controllers $C$ in $M$.
3:    **for** every controller $C$ in $Cs$ **do**
4:       $C/M := SubMPTS(M, C)$; //Get the controlled system $C/M$.
5:       $G := ProductAutomaton(C/M, R)$; //Build the product automaton $G$.
6:       $ACs := ComputeAllAC(G)$; //Compute the accepting component set $ACs$.
7:       $p := ModelChecking(G, ACs)$; //The value of formula $P_{=?}F(ACs)$ is computed using the PCTL probabilistic model checking.
8:       **if** $p \geq \vartheta$ **then**
9:          **return** $C$;
10:      **end if**
11:    **end for**
12:    **return** null;

For the specific details of translating LTL formulas into DRAs and PCTL probability model checking in Algorithm 1, readers are referred to the literature [14][15], and the paper will not elaborate further. The process of controller enumeration is relatively straightforward and thus will not be discussed in detail here. Subsequently, this paper will provide a detailed introduction to the acquisition of the controlled system, the construction of the product automaton, and the computation of the accepting component set.

### 4.2. Controlled System

Essentially, the controlled system $C/M$ is a sub model of the system model $M$, where the transitions between states in the controlled system are determined by the controller $C$. Referring to the definition of the controlled system in Section 3.3, Algorithm 2 provides a method for obtaining the controlled system under a given controller $C$ from the system model $M$.

**Algorithm 2**: *SubMPTS*($M, C$)

**Input**: System Model $M = <\Sigma, X, A, \{p_1, ..., p_n\}, p, f, x_0, AP, L>$, Controller $C$.

**Output**: Controlled System
$C/M = <\Sigma^{C/M}, X^{C/M}, A^{C/M}, \{p_1, ..., p_n\}^{C/M}, p^{C/M}, f^{C/M}, x_0^{C/M}, AP^{C/M}, L^{C/M}>$.

1:    $\Sigma^{C/M} := \Sigma, X^{C/M} := X, A^{C/M} := A, \{p_1, ..., p_n\}^{C/M} := \{p_1, ..., p_n\}, x_0^{C/M} := x_0, AP^{C/M} := AP, L^{C/M} := L$;
2:    **for** every state $x$ in $X^{C/M}$ **do**
3:       $\vec{A}(x) := \{<\vec{a^P}, \vec{a^Q}> | \vec{a^P} = C(x), \vec{a^Q} \in \vec{A^Q}(x)\}$; // $\vec{A}(x)$: the set of joint actions that can be enabled under the control of controller $C$ in state $x$; $\vec{A^Q}(x)$: the set of joint actions that can be enabled by players in $\Sigma^Q$ in state $x$.
4:       **for** every action $\vec{a} = <\vec{a^P}, \vec{a^Q}> = <a_1, ..., a_k>$ in $\vec{A}(x)$ **do**
5:          $p^{C/M}(x, \vec{a}) := \prod_{i \in \Sigma^Q} p_i(x, a_i)$;

6:       $f^{C/M}(x,\vec{a}) := f(x,\vec{a})$;
7:   **end for**
8: **end for**
9: **return** $C/M :=< \Sigma^{C/M}, X^{C/M}, A^{C/M}, \{p_1,...,p_n\}^{C/M}, p^{C/M}, f^{C/M}, x_0^{C/M}, AP^{C/M}, L^{C/M} >$;

### 4.3. Product Automaton

This section formally presents the construction rules for the product automaton. Given an MPTS $C/M =< \Sigma^{C/M}, X^{C/M}, A^{C/M}, \{p_1,...,p_n\}^{C/M}, p^{C/M}, f^{C/M}, x_0^{C/M}, AP, L^{C/M} >$ and a DRA $R =< X^R, 2^{AP}, f^R, x_0^R, Acc^R >$, their product automaton can be represented by a six-tuple, i.e.,

$$G = C/M \otimes R =< X, A, f, x_0, Acc, p >$$

where

(1) $X = X^{C/M} \times X^R$ is the state set.

(2) $A = A^{C/M}$ is the action set.

(3) $f : X \times \vec{A} \to X$ is the transition function. For any $(x,\vec{a}) = ((x^{C/M}, x^R), \vec{a}) \in X \times \vec{A}$, having $f(x,\vec{a}) = (f^M(x^M, \vec{a}), f^R(x^R, L^M(x^M)))$, where $x^{C/M} \in X^{C/M}$ and $x^R \in X^R$.

(4) $x_0 = (x_0^{C/M}, x_0^R)$ is the initial state.

(5) $Acc = \{(L_1, K_1), (L_2, K_2),...,(L_d, K_d)\}$ is the Rabin acceptance condition. For any $1 \le i \le d$, having $L_i = X^{C/M} \times L_i^R$ and $K_i = X^{C/M} \times K_i^R$.

(6) $p : X \times \vec{A} \to [0,1]$ is the state action probability function. For any $(x,\vec{a}) = ((x^{C/M}, x^R), \vec{a}) \in X \times \vec{A}$, having $p(x,\vec{a}) = p^{C/M}(x^{C/M}, \vec{a})$.

It can be observed that the action set of $G$ is consistent with that of $C/M$. For an infinite path $\rho = (x_0^{C/M}, x_0^R)\vec{a_0}(x_1^{C/M}, x_1^R)...$ in $G$, if there exists $1 \le i \le d$ such that $Inf(\rho) \cap L_i = \emptyset \wedge Inf(\rho) \cap K_i \ne \emptyset$, then the infinite path is said to be accepted by $G$. Every accepted path $\rho = (x_0^{C/M}, x_0^R)\vec{a_0}(x_1^{C/M}, x_1^R)...$ in $G$ can be mapped to a run $\rho^{C/M} = x_0^{C/M}\vec{a_0}x_1^{C/M}...$ of $C/M$ and a run $\rho^R = x_0^R L^{C/M}(x_0^{C/M})x_1^R...$ of $R$, and the infinite word $\omega = L^{C/M}(x_0^{C/M})L^{C/M}(x_1^{C/M})... \in (2^{AP})^\infty$ is accepted by $R$, and vice versa. Clearly, $G$ captures the behaviour of both $C/M$ and $R$. According to the construction rules of the product automaton and the content of Section 1.3 and Section 2.1, the following relationship can be derived, i.e.,

$$Pr^{C/M}\{\rho^{C/M} \in Runs(C/M) \mid C/M, \rho^{C/M} \vDash \varphi\}$$
$$\Leftrightarrow Pr^{C/M}\{\rho^{C/M} \in Runs(C/M) \mid L(\rho^{C/M}) \vDash \varphi\}$$
$$\Leftrightarrow Pr^{C/M}\{\rho^{C/M} \in Runs(C/M) \mid L(\rho^{C/M}) \in \mathcal{L}(R_\varphi)\}$$
$$\Leftrightarrow Pr^G\{\rho^G \in Runs(G) \mid \rho^G \text{ is an accpeting path}\}$$

This is primarily because the product automaton extends $C/M$ by tracking the runs of $R$. For the derivation of this equivalence, one can refer to [16][985-989]. Based on the above, we can transform Problem 1 into the problem of computing the probability of accepting paths in the product automaton.

### 4.4. Accepting Component

To compute the probability of accepting paths in the product automaton, inspired by the LTL probability model checking [15], we extend the concept of strong connected components to adapt to the research presented study.

An accepting component of the product automaton $G = <X, A, f, x_0, Acc, p>$ is a two-tuple $AC = <S, P>$, where $S$ is a subset of $X$, i.e., $S \subseteq X$, and $P: S \times S \to [0,1]$ is the probability transition function, satisfying the following conditions:

(1) $<S, P>$ determines a sub model of $G$, meaning for any $s \in S$ and $\vec{a} \in \vec{A}(s)$, having $f(s, \vec{a}) \in S$. And for any $s$, $s' \in S$, having $P(s, s') = \sum_{\vec{a} \in \vec{A}(s,s')} p(s, \vec{a})$, where $\vec{A}(s, s') = \{\vec{a} \mid f(s, \vec{a}) = s'\}$ denotes the set of joint actions that can be enabled in state $s$ to transition to state $s'$. Additionally, for any $s \in S$, having $\sum_{s' \in S} P(s, s') = 1$.

(2) The graph formed by $<S, P>$ is strongly connected.

(3) There exists $(L, K) \in Acc$ such that for all $s \in S$ satisfying $s \notin L$ and there exists $s \in S$ satisfying $s \in K$.

Based on the definition of accepting paths and accepting component in the product automaton, we can further deduce the following: Let $ACs \subseteq X$ be the accepting component set, $x \in ACs$ if and only if $x$ appears in some accepting component of $G$, then

$$Pr^G\{\rho^G \in Runs(G) \mid \rho^G \text{ is an accpeting path}\}$$
$$\Leftrightarrow Pr^G\{\rho^G \in Runs(G) \mid \rho^G \text{ contains a state in } ACs\}$$

The derivation of this equivalence can be found in [16][985-989]. Thus, we transform the problem of computing the probability of satisfying the LTL formula $\varphi$ under controller $C$ in system model $M$ into (i) the computation of accepting component set $ACs$ in product automaton $G$, and (ii) the computation of the probability of reaching accepting component set $ACs$ from initial state of product automaton $G$.

Algorithm 3 provides a method for computing the acceptance component set. This algorithm first uses the Tarjan algorithm to compute the strong connected components, ensuring that the acceptance component meets the second requirement of the acceptance component definition. Then, it determines whether each strong connected component is a bottom strongly connected component, ensuring that the acceptance component meets the first requirement of the acceptance component definition. Finally, it checks whether the states in each bottom strongly connected component satisfy some Rabin acceptance conditions, ensuring that the acceptance component meets the third requirement of the acceptance component definition. The acceptance component set is composed of all states in the acceptance components obtained from the above process.

**Algorithm 3**: *ComputeAllAC(G)*

**Input**: Product Automaton $G = <X, A, f, x_0, Acc, p>$

**Output**: Accepting Component $ACs$

1: $ACs := \emptyset, BSCCs := \emptyset, SCCs := \emptyset$ ; //Initialize the accepting component set $ACs$, the bottom strongly connected component set $BSCCs$, and the strong connected component set $SCCs$.

2: $SCCs := Tarjan(G)$ ; //Use the Tarjan algorithm to compute all the strong connected components in $G$.

3: **for** every $SCC$ in $SCCs$ **do**

4:     $flag1 := 1$; //A temporary variable that identifies whether the $SCC$ is or not $BSCC$.

5:     **for** every state $x$ in $SCC$ **do**

6:         **for** every action $\vec{a}$ in $\vec{A}(s)$ **do**

7:             **if** $f(x, \vec{a}) \in SCC$ **then**

```
 8:            continue;
 9:         else
10:            flag1 := 0;
11:            break;
12:         end if
13:      end for
14:    end for
15:    if flag1 = 1 then
16:       BSCCs := BSCCs ∪ SCC;
17:    end if
18: end for
19: for every BSCC in BSCCs do
20:    flag2 := 0; // A temporary variable that identifies whether the BSCC is or not AC.
21:    for every (L, K) in Acc do
22:       if BSCC ∩ L = ∅ and BSCC ∩ K ≠ ∅ then
23:          flag2 := 1;
24:          break;
25:       else
26:          continue;
27:       end if
28:    end for
29:    if flag2 := 1 then
30:       ACs := ACs ∪ BSCC;
31:    end if
32: end for
33: return ACs;
```

## 5. CASE STUDY

This section uses the classic philosopher dining system as an example to illustrate the effectiveness of the proposed method. In this system, there are three philosophers, denoted as $P_1$, $P_2$ and $P_3$. For each philosopher, the possible states include thinking $A$, waiting for the left fork $B$ (i.e., not yet holding the left fork), waiting for the right fork $C$ (i.e., already holding the left fork), waiting for the right fork $D$ (i.e., already holding the left fork), and eating $E$. The possible actions include thinking $a$, becoming hungry $b$, picking up the left fork $c$, picking up the right fork $d$, putting down the left and right fork $e$, and waiting for the fork $f$. It is assumed that once a philosopher obtains both forks, he immediately starts eating, and after completing the meal, he simultaneously put back both forks. Figures 1, 2, and 3 respectively illustrate the probabilistic transition models for the three philosophers.

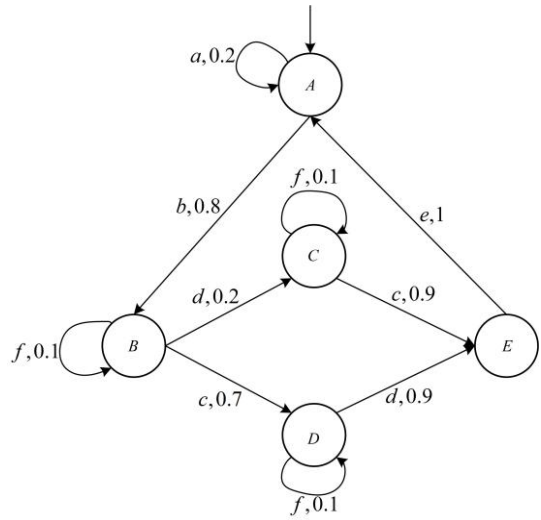

Fig.1. Probabilistic transition model of philosopher $P_1$

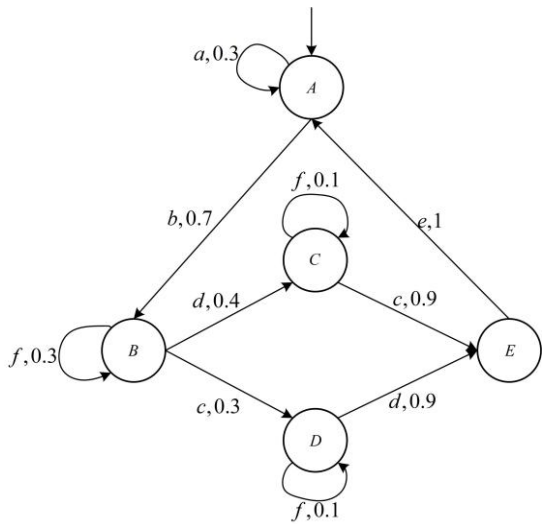

Fig.2. Probabilistic transition model of philosopher $P_2$

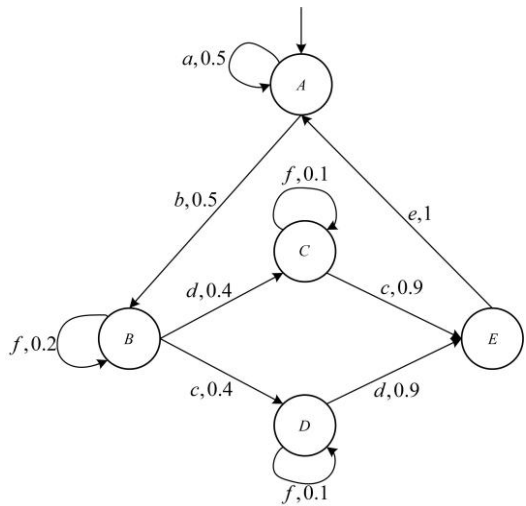

Fig.3. Probabilistic transition model of philosopher $P_3$

Combining the individual probabilistic transition models of the three philosophers yields a multi-agent probabilistic system model, which can be represented using the multi-agent probabilistic transition system proposed in Section 3.1. The details are as follows:

(1) The set of players is $\Sigma = \{1,2,3\}$, where "1" represents philosopher $P_1$, "2" represents philosopher $P_2$, and "3" represents philosopher $P_3$. In this case, philosophers $P_1$ and $P_2$ are grouped into an alliance, hence $\Sigma^P = \{1,2\}$, $\Sigma^Q = \{3\}$.

(2) The state set is $X = X_1 \times X_2 \times X_3$, where $X_1$, $X_2$ and $X_3$ represent the state sets of philosophers $P_1$, $P_2$, and $P_3$, respectively, and $X_1 = X_2 = X_3 = \{A,B,C,D,E\}$.

(3) The action set is $A = \{a,b,c,d,e,f\}$. The joint action set is $\vec{A} = \overrightarrow{A^P} \times \overrightarrow{A^Q} = (A_1 \times A_2) \times A_3$, where $A_1 = A_2 = A_3 = A$, $\overrightarrow{A^P} = A_1 \times A_2$, $\overrightarrow{A^Q} = A_3$.

(4) The probability distribution set is $\{p_1, p_2, p_3\}$. $p_1$, $p_2$ and $p_3$ can be obtained from Figures 1, 2, and 3, respectively. For instance, in state $x = (B,A,A) \in X$, the probability of philosopher $P_1$ executing action $c$ is $p_1(x,c) = p_1(B,c) = 0.7$, the probability of executing action $d$ is $p_1(x,d) = p_1(B,d) = 0.2$, and the probability of executing action $f$ is $p_1(x,f) = p_1(B,f) = 0.1$, satisfying $\sum_{a \in A} p_1(x,a) = 1$. The remaining cases can be obtained similarly.

(5) The probability function $p$ can be obtained as follows: Taking the state $x = (B,A,A) \in X$ and the joint action $\vec{a} = <c,b,b>$ as an example, the probability of executing the joint action $\vec{a}$ in state $x$ is $p(x,\vec{a}) = p_1(x,c) \cdot p_2(x,b) \cdot p_3(x,b) = p_1(B,c) \cdot p_2(A,b) \cdot p_3(A,b) = 0.7 \times 0.7 \times 0.5 = 0.245$. The remaining cases can be obtained similarly.

(6) The transition function $f$ can be obtained as follows: Similarly, taking the state $x = (B,A,A) \in X$ and the joint action $\vec{a} = <c,b,b>$ as an example, assuming $f_1$, $f_2$, and $f_3$ are the transition functions from Figures 1, 2, and 3, respectively, then $f(x,\vec{a}) = (f_1(B,c), f_2(A,b), f_3(A,b)) = (D,B,B)$. The remaining cases can be obtained similarly.

(7) The initial state is $x_0 = (A,A,A)$.

(8) The set of atomic propositions is $AP = \{q_1, q_2, q_3, q_4\}$, where proposition $q_1$ indicates "philosopher $P_1$ is eating", $q_2$ indicates "philosopher $P_2$ is eating", $q_3$ indicates "philosopher $P_3$ is eating", and $q_4$ indicates "no philosopher is eating".

(9) The proposition on each state are as follows: For states $x \in \{E\} \times \{A,B,C,D\} \times \{A,B,C,D\}$, having $L(x) = \{q_1\}$; for states $x \in \{A,B,C,D\} \times \{E\} \times \{A,B,C,D\}$, having $L(x) = \{q_2\}$; for states $x \in \{A,B,C,D\} \times \{A,B,C,D\} \times \{E\}$, having $L(x) = \{q_3\}$; for states $x \in \{A,B,C,D\} \times \{A,B,C,D\} \times \{A,B,C,D\}$, having $L(x) = \{q_4\}$.

For this case, this paper considers the following property specification:
$$\varphi := G(Fq_1 \wedge Fq_2)$$

The intuitive meaning of this specification is that philosophers 1 and 2 will always eventually be in the dining state. The DRA obtained from this specification is shown in Figure 4. The Rabin acceptance condition of this automaton is $Acc = \{(L_1 = \varnothing, K_1 = \{s_1\})\}$.

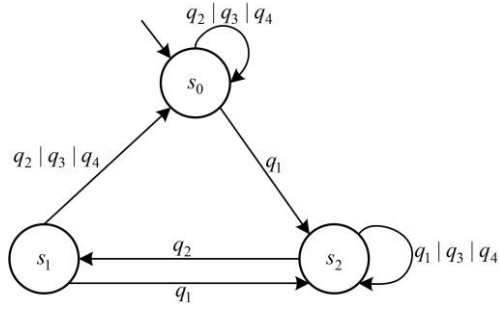

Fig.4. DRA transformed from $\varphi := G(Fq_1 \wedge Fq_2)$

Using the algorithm presented in Section 4 to solve for the controller, the obtained controller that meets the requirements is shown in Table 1. In this case, the LTL formula is converted into a DRA using the tool LTL2DSTAR[17], and the tool PRISM[18] is used to perform probabilistic model checking, with the probability threshold $\vartheta$ set to 0.8. After determining the strategies for some states, it is possible to identify some states as unreachable, hence only the strategy choices for reachable states are listed in Table 1, and the strategy choices for unreachable states have no impact on the satisfaction of the system model's probabilistic properties. The controlled system model under the action of this controller is shown in Figure 5. After verification by the PRISM tool, the probability that the controlled system satisfies the LTL formula $G(Fq_1 \wedge Fq_2)$ is 1.

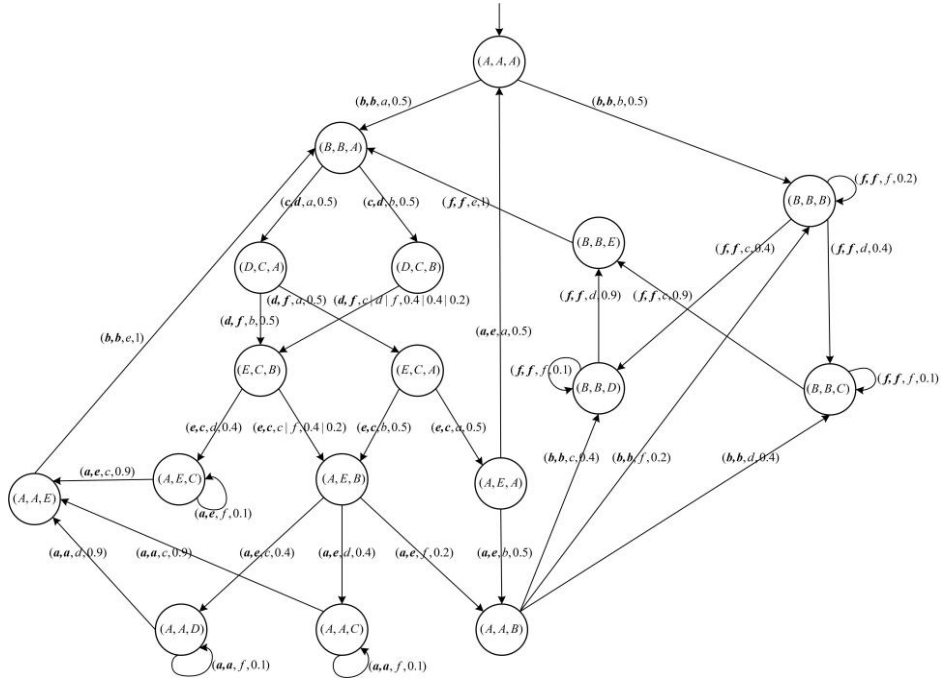

Fig.5. The controlled system under the action of controller

Table 1. The results of controller synthesis

| Number | State   | Joint action | Number | State   | Joint action |
|--------|---------|--------------|--------|---------|--------------|
| 1      | (A,A,A) | (b,b)        | 10     | (E,C,B) | (e,c)        |
| 2      | (B,B,A) | (c,d)        | 11     | (A,E,A) | (a,e)        |
| 3      | (B,B,B) | (f,f)        | 12     | (A,E,B) | (a,e)        |

| 4 | (B,B,D) | (f,f) | 13 | (A,E,C) | (a,e) |
| 5 | (D,C,A) | (d,f) | 14 | (A,A,B) | (b,b) |
| 6 | (D,C,B) | (d,f) | 15 | (A,A,C) | (a,a) |
| 7 | (B,B,C) | (f,f) | 16 | (A,A,D) | (a,a) |
| 8 | (E,C,A) | (e,c) | 17 | (A,A,E) | (b,b) |
| 9 | (B,B,E) | (f,f) | | | |

## 6. CONCLUSIONS AND FUTURE WORK

Controller synthesis is a core research topic in the field of discrete event systems and holds significant research value. This paper focuses on semi-cooperative semi-competitive multi-agent probabilistic discrete event systems and proposes a controller synthesis method based on LTL specifications. By constructing product automaton and computing accepting component, a controller synthesis algorithm based on probabilistic model checking is designed. Finally, the effectiveness of the proposed method is verified through a philosopher dining case. Unlike the studies in literatures [2-9], this paper targets semi-cooperative semi-competitive multi-agent probabilistic discrete event systems, which have broader application value to some extent compared to single-agent discrete event system research. Additionally, unlike the studies in literatures [2-5] that only use formal languages or probabilistic languages to describe control specifications, this paper combines LTL specifications and probabilities, allowing us to consider both the temporal and quantitative properties of systems that satisfy given specifications simultaneously. Furthermore, this paper proposes a controller synthesis method based on probabilistic model checking, which differs from the methods in literatures [6-9] that use reactive synthesis and bounded synthesis techniques for controller synthesis. Our method establishes a connection between controller synthesis research and model checking research. However, for the method proposed in this paper, when the number of agents in the system continues to increase, the state space of the entire system will experience explosive growth. Therefore, whether methods that alleviate the state space explosion in model checking can be introduced into controller synthesis is a problem worth considering in future work.

## ACKNOWLEDGEMENTS

This work was supported by the Fundamental Research Funds for the Central Universities under Grant NJ2024030.

## AUTHORS

**Zining Cao** received the Ph.D. degree in 2001. He is now a professor in the College of Computer Science and Technology at Nanjing University of Aeronautics and Astronautics. His current research interests include formal methods in software engineering and logic in computer science. Zining Cao is the corresponding author. Email: caozn@nuaa.edu.cn